# Suppression of tunneling two-level systems in ultrastable glasses of indomethacin

Tomás Pérez-Castañeda[a], Cristian Rodríguez-Tinoco[b], Javier Rodríguez-Viejo[b,c] & Miguel A. Ramos[a,1]

[a]Laboratorio de Bajas Temperaturas, Departamento de Física de la Materia Condensada, Condensed Matter Physics Center (IFIMAC) and Instituto Nicolás Cabrera, Universidad Autónoma de Madrid, E-28049 Madrid, Spain.

[b]Group of Nanomaterials and Microsystems. Department of Physics, Universitat Autònoma de Barcelona, Bellaterra 08193, Spain.

[c]MATGAS Research Center, Campus UAB, 08193 Bellaterra, Spain.

[1]To whom correspondence should be addressed. E-mail: miguel.ramos@uam.es

**Keywords:** tunneling two-level systems; specific heat; glass transition; organic glasses

# ABSTRACT

Glasses and other non-crystalline solids exhibit thermal and acoustic properties at low temperatures anomalously different from those found in crystalline solids, and with a remarkable degree of universality. Below a few K, these universal properties have been successfully interpreted using the Tunneling Model, which has enjoyed (almost) unanimous recognition for decades. Here we present low-temperature specific-heat measurements of ultrastable glasses of indomethacin that clearly show the disappearance of the ubiquitous linear contribution traditionally ascribed to the existence of tunneling two-level systems (TLS). When the ultrastable thin-film sample is thermally converted into a conventional glass, the material recovers a typical amount of TLS. This remarkable suppression of the TLS found in ultrastable glasses of indomethacin is argued to be due to their particular anisotropic and layered character, which strongly influences the *dynamical* network and may hinder isotropic interactions among low-energy defects, rather than to the *thermodynamic* stabilization itself. This explanation may lend support to the criticisms by Leggett and others to the standard Tunneling Model, although more experiments in different kinds of ultrastable glasses are needed to ascertain this hypothesis.

**Introduction**

Glasses or amorphous solids are well known (1,2) to exhibit thermal and acoustic properties very different from those of their crystalline counterparts. Even more strikingly, many of these properties are very similar for any glass, irrespective of the type of material, chemical bonding, etc. Hence the low-temperature properties of non-crystalline solids are said to exhibit a *universal* "glassy behavior". In particular, below 1−2 K the specific heat of glasses depends approximately linearly on temperature, $C_p \propto T$, and the thermal conductivity almost quadratically, $\kappa \propto T^2$, in clear contrast with the cubic dependences successfully predicted by Debye theory for crystals. In addition, a broad maximum in $C_p/T^3$ (originated from the so-called "boson peak" in the reduced vibrational density of states $g(\omega)/\omega^2$) is also typically observed in glasses around 3−10 K, as well as a universal plateau in the thermal conductivity $\kappa(T)$ in the same temperature range (1,2).

Very soon after the seminal paper by Zeller and Pohl (1) in 1971, Phillips (3) and Anderson *et al*. (4) independently introduced the well-known standard Tunneling Model (TM). The fundamental idea of the TM is the ubiquitous existence of atoms or small groups of atoms in amorphous solids due to the intrinsic atomic disorder, which can perform quantum tunneling between two configurations of very similar energy, usually named two-level systems (TLS). This simple model was able to account for the abovementioned thermal and acoustic *anomalies* of glasses below 1−2 K, and soon acquired unanimous recognition. Only very few authors (5) posed then criticisms against the standard TM, pointing out how improbable was that a random ensemble of independent tunneling states would produce essentially the same universal constant for the thermal conductivity or the acoustic attenuation in any substance. Indeed, significant discrepancies with the TM below ~100mK have also been reported (6-9), in particular a strong, unexpected strain dependence in the acoustic properties exhibited by both dielectric and metallic glasses (8,9). More recently, single-molecule spectroscopy experiments above 2 K have shown that the spectral dynamics in low-molecular-weight glasses and short-chain polymers on a microscopic scale cannot be described within the standard TM, unlike the single-molecule spectral dynamics in long-chain polymers (10).

Conventional glasses are obtained by cooling fast enough the liquid. Slowing down the cooling rate drives the system to lower energy positions in the potential energy landscape, being the minimum cooling rate ruled by the occurrence of crystallization. This limitation has been recently defeated by growing glasses directly from the vapor-phase (11,12). Those glasses, dubbed ultrastable glasses, can be synthesized by physical vapor-deposition in short-time scales and show unprecedented thermodynamic and kinetic stability (13−17). Changing the growth parameters has a strong effect on the properties of the glass. An appropriate deposition rate in combination with an optimal substrate temperature (typically around 0.85 $T_g$, where $T_g$ stands for the glass-transition temperature) have proved to drastically favour two-dimensional mobility and, as a consequence, access to local minima of very low energy in the potential-energy landscape (18,19). An ordinary glass obtained by supercooling the liquid should theoretically be aged for $10^3−10^9$ years in order to achieve the same stability and density of these vapor-deposited glasses (20).

To the best of our knowledge, no study has been performed up to date to investigate the possible effects that the dramatic increase in thermodynamic and kinetic stability of vapor-deposited ultrastable glasses could have on the universal low-temperature anomalies of glasses. In what follows we present low-temperature specific heat data for ultrastable glasses prepared by physical vapor deposition. In particular, we measured ultrastable, ordinary (cooled at 10 K/min from the liquid) and crystalline samples of indomethacin in the range 0.18 K ≤ T ≤ 32 K. We also describe the effect of the loss of stability and water absorption on the low-*T* properties of an ultrastable glass, providing complementary enlightening information about the nature of the TLS suppression.

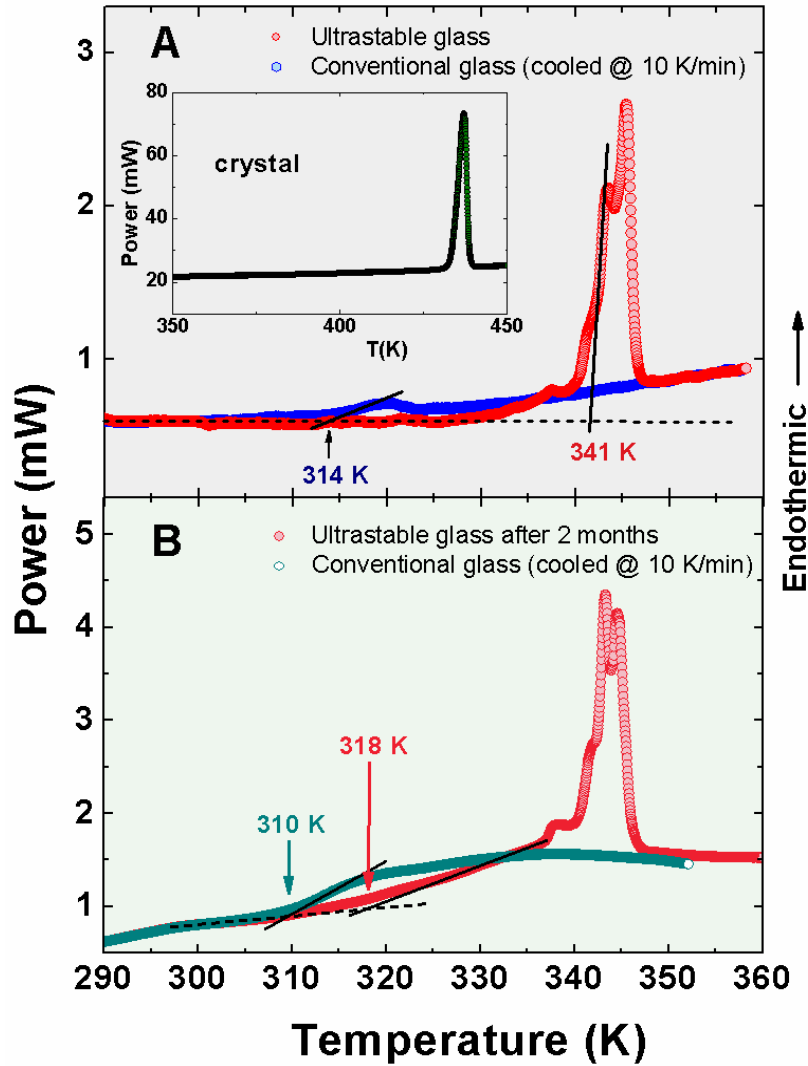

**Fig. 1.** *(A)* Calorimetric up-scans for conventional (cooled at a rate of 10 K/min) and ultrastable glasses of indomethacin. Inset shows the melting of the crystal at $T_m = 430$ K. The glass transition temperatures of the conventional and ultrastable glasses are determined by the onset in the heat flow curves, $T_g^{conv} \equiv T_{on}^{conv} = 314$ K and $T_{on}^{USG} = 341$ K, depicted by the intersection of the corresponding solid line (taking the main peak for the USG) and the extrapolated dashed line of the glass. *(B)* Calorimetric curves for a 50-μm-thick ultrastable glass after being stored in poor vacuum conditions for 2 months, and for the subsequent conventional glass obtained by cooling the melt sample at 10 K/min. For these degraded (including water absorption) glasses, $T_{on}^{conv} = 310$ K and $T_{on}^{USG} = 318$ K, here signaling the onset of the deviation from the glass background (dashed line).

## Results

We prepared indomethacin (IMC) samples by vapor-deposition at 0.85 $T_g$ with a growth rate of 0.15±0.05 nm/s on Si(100) substrates for low-temperature specific heat and x-ray diffraction (XRD) measurements, as well as on Al pans for Differential Scanning Calorimetry (DSC) analysis. In this way, we produced samples with thicknesses ranging 50−80 μm with the aim to maximize the signal-to-noise ratio required in the specific-heat measurements (see Materials and Methods). The extraordinary kinetic and thermodynamic stability of the ultrastable glass (USG) is characterized by the noteworthy difference in the calorimetric onset temperature $T_{on}$ for the glass transition with respect to the conventional (conv) glass cooled at 10 K/min ($\Delta T_{on} = T_{on}^{USG} - T_{on}^{conv} = 27$ K), as shown in Fig. 1A, as well as in the limiting fictive temperature ($\Delta T_f = T_f^{conv} - T_f^{USG} = 33$ K) obtained from the corresponding enthalpy extrapolations. These data agree well with previous results found in the literature (11,16). The melting curve of the crystalline phase is also shown in the inset for comparison. The presence of several peaks in the calorimetric curve of the ultrastable glass is related to changes of stability, produced by variations in the growth rate during deposition (20).

In Fig. 1B, we show DSC measurements for a USG sample, after being stored in poor vacuum conditions for two months. This causes a loss of stability and water absorption (21) which shifts 4 K downwards the calorimetric glass transition of the conventional glass. The calorimetric data shows that the sample has notably decreased its thermodynamic and kinetic stability, since a significant fraction of the glass exhibits a $\Delta T_{on}$ of only 8 K, relative to the conventional glass.

The specific heat of several indomethacin samples in different states was measured in the temperature range 0.18 K ≤ $T$ ≤ 32 K. Fig. 2A shows the whole specific-heat data in the Debye-reduced $C_P/T^3$ representation. Fig. 2B amplifies the

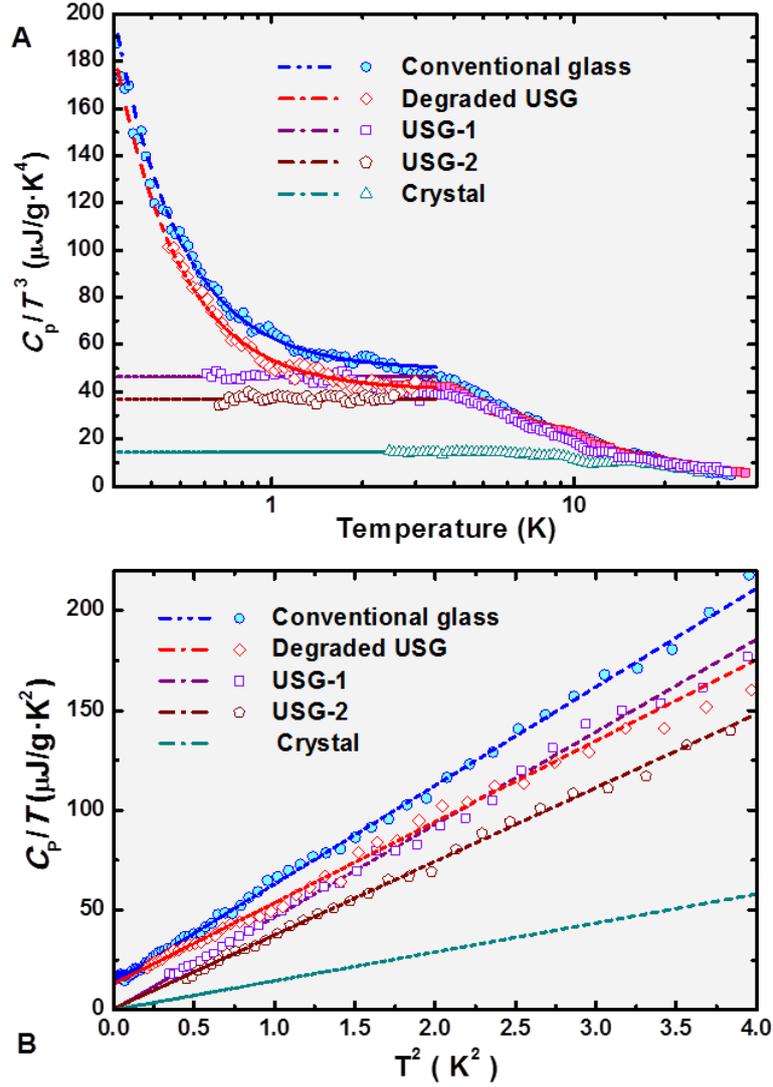

**Fig. 2.** Specific-heat data for ultrastable glasses of indomethacin 50 µm- (USG-1) and 80µm- (USG-2) thin films, compared to the crystalline phase (Debye extrapolated at lower temperatures) and the conventional glass. A degraded ultrastable glass (see text) has also been measured and is presented. Dashed lines show the corresponding linear fits $C_P = c_{TLS} \cdot T + c_D \cdot T^3$ for experimental data below 2 K. (*A*) Debye-reduced $C_P/T^3$ versus $T$ representation; (*B*) $C_P/T$ versus $T^2$ plot at very low temperatures to determine the TLS and the Debye coefficients, which are given in Table 1.

very-low-temperature region in the usual $C_P/T$ vs $T^2$ plot where a least-squares linear fit provides the TLS linear term (the intercept with the *y* axis) and the Debye coefficient (the slope). The crystal of indomethacin exhibits the expected $C_P \propto T^3$ below 8 K. The same shoulder-like behavior (a very shallow *boson peak*, as typically occur in other fragile glass formers (22)) is observed in both ultrastable and ordinary glasses below ~5

K in Fig. 2A. This is consistent with earlier Raman-scattering experiments in indomethacin (23), where a hardly visible boson peak was found both in the normal glass state and in a high-pressure amorphous state. As can also be seen in Fig. 2, a modest difference between the Debye levels of the two ultrastable samples (USG-1 and USG-2, prepared with slightly different conditions) is observed, both of them below that of the conventional glass. It is noteworthy that the Debye coefficient of the crystal is very much smaller (its elastic constants are much harder) than those of the glasses. The so-obtained calorimetric Debye coefficients agree very well with the elastic Debye coefficients obtained from room-temperature sound velocity and mass density data from the literature (24), as can be observed in Table 1. This agreement further supports our analysis of the low-temperature specific heat curves.

**Table 1. Specific heat coefficients**

| Sample state | $c_{TLS}$ ($\mu$J/g·K$^2$) | $c_D$ ($\mu$J/g·K$^4$) | $c_D^{elas}$ ($\mu$J/g·K$^4$) |
|---|---|---|---|
| CRYSTAL | – | 15.0 ±0.3 | – |
| CONVENTIONAL GLASS | 13.7 ±0.3 | 49.4 ±0.2 | 51 |
| ULTRASTABLE GLASS #1 | 0.2 ±0.9 | 46.4 ±0.6 | 41 |
| ULTRASTABLE GLASS #2 | 0.02 ±0.8 | 36.9 ±0.4 | 41 |
| DEGRADED USG | 13.0 ±0.7 | 40.6 ±0.5 | – |

**Table 1.** Coefficients and statistical errors from the least-squares linear fits at low temperatures to the function $C_P = c_{TLS} \cdot T + c_D \cdot T^3$ (see Fig. 2B). The last column indicates the expected Debye coefficient $c_D^{elas}$ for conventional and ultrastable glasses of IMC, obtained from published elastic data at room temperature (24).

Nevertheless, the most surprising behavior found in both ultrastable glasses is the full suppression of the linear term of the specific heat ascribed to the tunneling TLS.

This is demonstrated in Fig. 2B, where the intercept with the ordinate axis goes to zero within experimental error (see Table 1), in clear contrast with the case of conventional glass, or even for a degraded ultrastable glass after being stored in poor vacuum for two months. We note that the lack of experimental points below 0.6 K for the ultrastable glasses (USG-1 and USG-2 in Fig. 2) is an experimental manifestation of the dramatic reduction in the specific heat at very low temperatures compared to the conventional glasses. In the former case, the total measured heat capacity becomes so low that it rapidly approaches the contribution of the addenda and hence the net specific heat of the sample cannot be assessed with accuracy.

Some glasses obtained by physical vapor deposition show evidence of molecular anisotropy which is partly due to the growth method of thin films from the vapor phase. In particular, ultrastable IMC glasses exhibit an extra, low-$q$, peak in wide-angle x-ray scattering (WAXS) spectra (25) and birrefrigence in ellipsometric measurements (26). Also, computer simulations of "stable" glasses of trehalose revealed a distinct layered structure along the direction normal to the substrate, that was absent in the "ordinary" glass (27). However, recent experiments conducted in vapor-deposited glasses of the four isomers of tris-naphthylbenzene (TNB) have shown that anisotropy is unrelated to glass stability, rather being a secondary feature that will appear more or less prominently depending upon molecular structure (28).

In Fig. 3 we show that the low-$q$ peak appears indeed in the WAXS pattern of our vapor-deposited ultrastable glass, whereas it is absent in the conventionally prepared glass. The presence of this peak for the ultrastable glass should be related to some sort of molecular order along the growth direction, perpendicular to the substrate, as clearly revealed in the in-plane/out-of plane diffraction experiments of Figure 3B. This orientation may be enabled by the high mobility of the indomethacin molecules when they impinge the substrate surface from the vapor (25). Molecular orientation in vapor-deposited glassy films of organic semiconductors has been widely recognized as a potential source to increase carrier mobility through an enhancement of π-conjugation. The longer the molecular length is, the larger the anisotropy of the molecular orientation becomes (29).

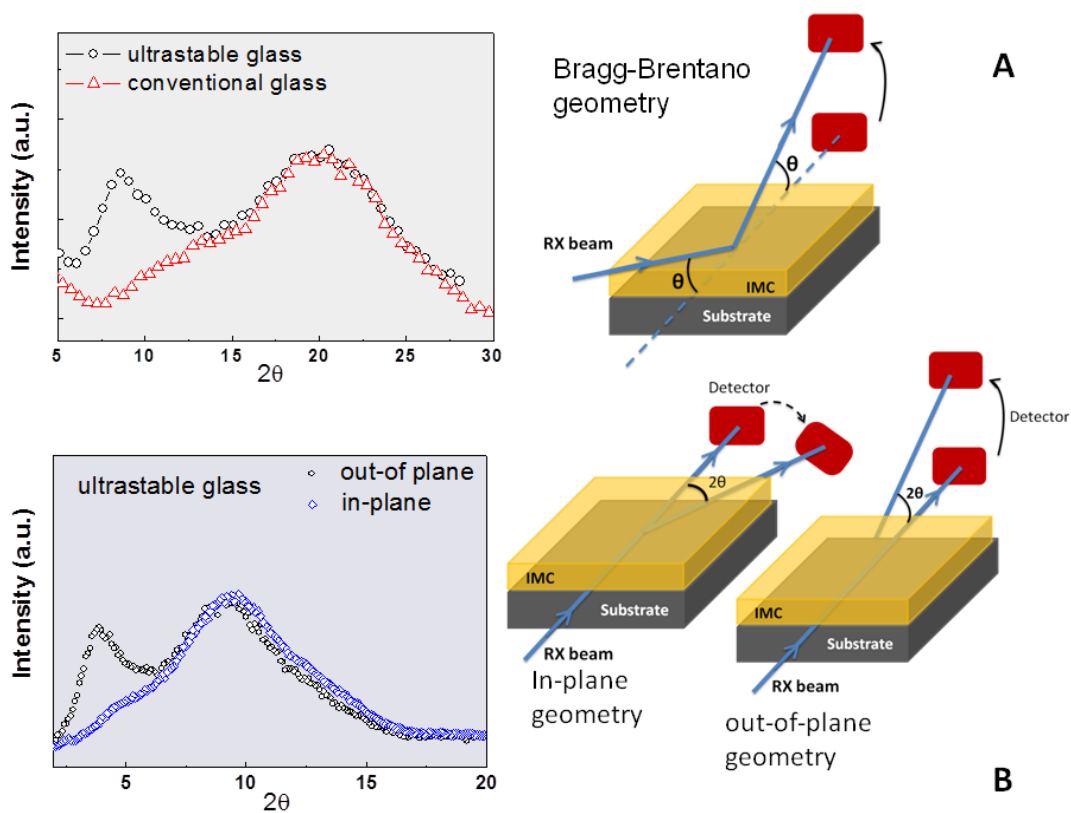

**Fig. 3.** (*A*) Wide-angle x-ray scattering spectra comparing ultrastable (USG) and conventional (conv) glasses of IMC. USG (conv) data are the result of box-averaging over 20 (10) points, respectively. The appearance of a peak at low-*q* angles in the USG is related to molecular ordering in the direction perpendicular to the substrate, an indication of molecular anisotropy. (*B*) In-plane and out-of-plane scattering experiments using synchrotron x rays on layered USG of IMC. The disappearance of the low-angle peak for the USG film in the in-plane experiment indicates that extra ordering occurs mainly in the z direction (perpendicular to the substrate) and is absent in the *xy* plane. The sketches show the configuration of the x-ray scattering experiments.

Let us discuss more specifically our case of indomethacin. In melt-quenched or grinded amorphous IMC the most favorable nearest neighbor packing direction occurs normal to the plane containing the indole ring, as occurs in the *γ* crystalline polymorph.

The dominant XRD peak at $2\theta\sim20^o$ (Cu K$_\alpha$, Fig. 3A) corresponds to an average distance between nearest neighbors of 0.45 nm, which matches the IMC molecular thickness when including the van der Waals radii. Interaction between molecules thus occurs from hydrogen-bonding cyclic dimers through the carboxilic groups. Nonetheless, vapor-deposited IMC glasses exhibit another strong XRD peak at $2\theta\sim8.5^o$ (Fig. 3A), indicative of an additional order within the structure, with a molecular packing distance of 1.1 nm. This value approximately corresponds to the distance between IMC molecules along the long-axis. As we have verified from in-plane and out-of-plane synchrotron XRD experiments (Fig. 3B), molecular anisotropy occurs mainly in the growth direction. This is again a strong indication of a layered growth (27).

## Discussion

Our experimental work is not the first one reporting lack of TLS in an amorphous solid. Nonetheless, previous reports (30–33) claiming the absence of TLS in amorphous solids are scarce and somewhat controversial. Angell *et al*. (34) proposed the designation of "superstrong liquids" for some "tetrahedral liquids" which could be potential "perfect glasses", with a residual entropy near zero, and where the defect-related boson peak and TLS excitations were weak or absent. Specifically, they identified two instances where TLS had been reported to be absent: (i) amorphous silicon (a-Si); (ii) low-density amorphous (LDA) water.

Let us stress, however, that Pohl and co-workers (30) did find TLS in pure a-Si (the ideal "superstrong liquid"). It was only in 1 at. % hydrogenated silicon where they observed a dramatic reduction of the internal friction plateau, that is proportional to the amount of TLS weighted by the TLS-phonon coupling energy. The suppression of TLS in hydrogenated a-Si was attributed by the authors (30) to a more compact fourfold coordination due to the passivation of the dangling bonds by the hydrogen, hence producing an overconstrained vibrational network. On the other hand, Hellman and coworkers (31,32) have reported *thermal* measurements suggesting a zero density of TLS in some a-Si thin films. Their conclusion was based on the variation of the specific heat *above* 2 K in a series of a-Si thin-film samples with different densities and preparation methods. In general, measurements in these and other tetrahedrally-bonded amorphous semiconductors have given conflicting results (32).

The case of amorphous water seems clearer in this respect. At low temperatures, there are (at least) two different amorphous states of water, HDA and LDA, associated with a high density liquid (HDL) and a low density liquid (LDL), respectively. The LDL of water has been found to be the strongest of all liquids known (35). Agladze and Sievers (33) reported no far-infrared resonant absorption by TLS in LDA ice at low temperature, whereas HDA ice exhibited the typical TLS response of other glasses.

At first glance, one might thus ascribe the found suppression of the tunneling TLS in ultrastable glasses of IMC to the extraordinary stability (either thermodynamic or structural) of these particular glasses, and its corresponding large reduction in enthalpy or entropy. This conclusion seems however to be at odds with our opposite recent finding in a related system: *hyperaged* glasses of geological amber. In pristine amber, a bulk isotropic glass which has been aging for millions of years, the amount of TLS has been found to remain constant after rejuvenation (36,37). Then, it seems reasonable to seek other plausible explanations for this unusual disappearance of the TLS in ultrastable indomethacin.

As outlined in the Introduction, several authors (5,6,38) have pointed out that, besides the unexplained universality of the TM fitting parameters, the model in its original form neglects the fact that as a result of interaction with the strain (phonon) field, the tunneling TLS must acquire a mutual interaction. It can be shown (5,6) that the effective interaction between two TLS separated by a distance $r$ is dipolar elastic and the interaction strength goes as $\sim g/r^3$, with $g = \gamma^2/\rho v^2$, where $\gamma$ is the TLS-phonon coupling constant, $\rho$ is the mass density and $v$ is the sound velocity of a given substance. An ensemble of independent, non-interacting TLS would not be possible nor could justify the observed *quantitative* universality. Instead, the observed astonishing universality could emerge as the general result of some renormalization process of (almost) any ensemble of defects or manybody energy levels and stress matrix elements, interacting through the usual bath of thermal phonons, implying the existence of some crossover length scale $r_0$ (roughly estimated to be about 1.3 nm) (5).

Therefore, we speculate that the picture of a spherical volume of size $r_0{}^3$ comprising an isotropic random distribution of structural defects (TLS) embedded in a 3D vibrational lattice, allowing the interaction between resonant defects via the acoustic-phonon bath, may fail in the case of these layered and anisotropic ultrastable glasses of IMC. We suggest that a possible interpretation of the found suppression of TLS in ultrastable IMC thin-film glasses grown at 0.85 $T_g$ could then be related to the modification of the molecular interaction in vapor-deposited USG films, through a decrease of free hydrogen bonds and an enhancement of π-π interactions between chlorophenyl rings. As studied by Dawson *et al.* (21), water uptake in IMC increases with the decrease in stability, so that loss of stability and increase of water absorption are concomitant processes. Water absorption seems to take place by occupying free sites of the IMC glass where water can hydrogen bond (21). Thus, without modifying the intrinsic structure of the layer, absorbed water molecules are able to bridge IMC molecules through hydrogen bonds, so feeding the interconnection of the dynamical network (water is known to be a good plasticizer), and hence recovering the *interacting*

TLS excitations. Therefore, the found suppression of the two-level-systems would not be related to the extraordinary stability of the glass, but rather to the particular molecular arrangement ruled by the deposition conditions in this ultrastable glass.

In our view, highly-stable "ideal glasses" can be associated with a negligible excess in *configurational* entropy, whereas non-crystalline solids devoid of low-energy excitations (TLS, boson peak...) could be somehow associated with a small *vibrational* entropy. Both features may be related sometimes, but they are not automatically interlinked. Hyperaged amber (36,37) seems to be a good counterexample. It is also true, nonetheless, that the case of this canonical glassy polymer, without any crystalline reference, is far from the other cases considered. All the latter have shown −or they have been predicted− to present polyamorphism and liquid-liquid transitions, which could facilitate the creation of "superstrong" ideal glasses devoid of low-energy glassy excitations, whereas the former may still have enough residual entropy and TLS despite its extraordinary thermodynamic stabilization. Though the fact that the density of TLS remains exactly the same in amber after rejuvenation casts doubts on this interpretation.

Following the line of reasoning used above, however, it may well be that in some (few) cases an isotropic "perfect" glass (related to a "superstrong liquid") is lacking the necessary low-energy structural defects (34), whereas in some other cases an anisotropic and layered ultrastabilized glassy structure hinders a normal interaction between those low-energy molecular excitations mediated by lattice vibrations. In both cases, the result could be the same: a lack of an effective amount of (renormalized) TLS.

## Conclusion

In conclusion, the main result of our work is the found suppression within experimental error of tunneling TLS in ultrastable glasses of IMC, in clear contrast to the usual behavior observed in conventionally prepared glasses, and even in the same samples of previously ultrastable glasses after losing stability and absorbing water. This important finding has revealed another remarkable exception to the universal behavior of glasses at low temperatures, and hence should shed light on the unclear microscopic nature of the so-called tunneling TLS. Although other explanations have been regarded, we believe that our finding in very anisotropic and layered ultrastable glasses of IMC may support the arguments by Leggett and others (5,6,38), which have claimed against the standard TM and have emphasized the critical role played by the coupling of the (isotropic) structural defects of the amorphous solid to the acoustic phonons, and the consequent impossibility of regarding these entities as independent, non-interacting excitations.

Albeit our interpretation of these remarkable results is not definitive at all, we expect that our work can pave the way to trigger new investigations on these issues, including similar experiments in *non-layered* ultrastable glasses.

Finally, we want to stress that our results should not be considered relevant only for organic ultrastable glasses. On the contrary, it is the fact that we have identified another significant exception to the universality of TLS in glasses what is important, for it provides an invaluable hint to unveil the mystery of the TLS mentioned above.

## Materials and Methods

**Sample growth.** Indomethacin ($C_{19}H_{16}ClNO_4$, $T_g$= 315 K and $T_m$ ($\gamma$ form)= 428 K) crystalline powders with 99% purity were purchased from Sigma-Aldrich. Ultrastable glassy films of IMC were grown by vapor deposition, both on Si(100) substrates and differential scanning calorimetry Al pans, at 0.85 $T_g$, i.e. $T_{dep}$=266 K. An effusion cell filled with IMC powder was heated to achieve the desired deposition rate, as measured by a quartz crystal microbalance. When this rate was attained, the shutter was removed to start deposition. The thickness of the films ranged from 50 to 80 μm, due to the importance to enhance the signal-to-noise ratio in the low-$T$ specific-heat measurements. The growth rate was 0.15±0.05 nm/s. Note that variations of the growth rate during the time required to grow a 50−80 μm thick layer, i.e. a week, account for the observation of several peaks in the calorimetric traces of Fig. 1. All samples were stored in vacuum-sealed bags with desiccant in a freezer to minimize ageing prior to the low-$T$ specific heat measurements. Low-temperature data of the ultrastable glasses in high vacuum was acquired few days after preparation, with the exception of a sample stored in those conditions for two months, named 'degraded' ultrastable glass.

**Differential Scanning Calorimetry.** A DSC Perkin Elmer 7 was used to monitor the power absorbed/released during heating scans at a rate of 10 K/min on IMC thin films with masses of the order of 8−11 mg. The first scan typically corresponds to an ultrastable glass, while the second one is characteristic of a conventional glass obtained by cooling the liquid at 10 K/min. The variation in onset temperatures and enthalpy overshoots are a clear indication of the much higher kinetic and thermodynamic stabilities of an ultrastable glass compared to a conventional one.

**X-ray diffraction.** To confirm the glassy nature of the as-grown samples we carried out XRD measurements using an X-Pert from Phillips in the Bragg-Brentano configuration with Cu $K_\alpha$ radiation. The samples were scanned in Bragg-Brentano geometry from $2\theta=2^{o}$ to $30^{0}$ with an angular step of $0.025^{o}$ $(0.05^{o})$ and time per point of 18 (12) s for the USG and conventional IMC glasses, respectively. The raw data is box-averaged every 20 (10) points to improve signal-to-noise ratio (Fig. 3A). We also conducted experiments at the beamline ID28 of the European Synchrotron Radiation Facility (ESRF). The energy of the x rays was set to 23.725 eV. Photons were detected with a photodiode with the sample aligned parallel (in-plane) or perpendicular (out-of-plane) to the detector axis, as schematically shown in Fig. 3B.

**Thermal-relaxation calorimetry at low temperatures.** The IMC samples used for the low-temperature specific-heat measurements were all grown on silicon substrates of dimension 12 x 12 mm$^2$, and with typical masses $m \approx 0.1$g, what enabled us the handling of the samples, as well as optimal attachment and thermal contact to the calorimetric cell. Due to the versatility of the low-temperature calorimeter employed, the same experimental setup (36) was used in both a $^4$He cryostat and a $^3$He-$^4$He dilution refrigerator, to cover a temperature range $0.18$ K $\leq T \leq 32$ K. The calorimeter was calibrated using a clean silicon substrate, in order to accurately determine the empty-cell contribution to the total heat capacity curves and hence obtain the indomethacin specific heat. The heat capacity at low temperatures was measured using the relaxation method (See Supporting Information for more details).

## Acknowledgements

This work was financially supported by the Spanish MINECO FIS2011-23488 and MAT2010-15202 projects. JRV also acknowledges funding from Generalitat de Catalunya through 2009SGR-1225. T.P.C. thanks financial support from the Spanish Ministry of Education through the FPU grant AP2008-00030 for his PhD thesis. We acknowledge ID28 from ESRF and Tullio Scopigno for assistance with the XRD measurements.

## References

1. Zeller RC, Pohl RO (1971) Thermal Conductivity and Specific Heat of Noncrystalline Solids. *Phys Rev B* 4:2029–2041.

2. Phillips WA (1981) *Amorphous Solids: Low-temperature Properties. Topics in Current Physics*, *Vol. 24* (Springer, Berlin).

3. Phillips WA (1972) Tunneling States in Amorphous Solids. *J Low Temp Phys* 7:351–360.

4. Anderson PW, Halperin BI, Varma CM (1972) Anomalous Low-temperature Thermal Properties of Glasses and Spin Glasses. *Philos Mag* 25:1–9.

5. Yu CC, Leggett AJ (1988) Low Temperature Properties of Amorphous Materials: Through a Glass Darkly. *Comments Cond. Mat. Phys.* 14:231−251.

6. Burin AL, Natelson D, Osheroff DD, Kagan Y (1998) *Tunnelling Systems in Amorphous and Crystalline Solids*, ed Esquinazi, P (Springer, Berlin) Ch. 3.

7. Classen J, Burkert T, Enss C, Hunklinger S (2000) Anomalous Frequency Dependence of the Internal Friction of Vitreous Silica. *Phys Rev Lett* 84:2176–2179.

8. Ramos MA, König R, Gaganidze E, Esquinazi P (2000) Acoustic properties of amorphous metals at very low temperatures: Applicability of the tunneling model. *Phys Rev B* 61:1059–1067.

9. König R, et al. (2002) Strain dependence of the acoustic properties of amorphous metals below 1 K: Evidence for the interaction between tunneling states. *Phys Rev B* 65:180201(R).

10. Eremchev IY, Vainer YG, Naumov AV, Kador L (2011) Low-temperature dynamics in amorphous polymers and low-molecular-weight glasses—what is the difference? *Phys Chem Chem Phys* 13**:**1843–1848.

11. Swallen SF, et al. (2007) Organic Glasses with Exceptional Thermodynamic and Kinetic Stability. *Science* 315:353−356.

12. Dawson KJ, Kearns KL, Yu L, Steffen W, Ediger MD (2009) Physical vapor deposition as a route to hidden amorphous states. *Proc Natl Acad Sci USA* 106:15165−15170.

13. León-Gutiérrez E, García G, Lopeandía AF, Clavaguera-Mora MT, Rodríguez-Viejo J (2010) Size Effects and Extraordinary Stability of Ultrathin Vapor Deposited Glassy Films of Toluene. *J Phys Chem Lett* 1:341−345.

14. Ramos SL, Oguni M, Ishii K, Nakayama H (2011) Character of Devitrification, Viewed from Enthalpic Paths, of the Vapor-Deposited Ethylbenzene Glasses. *J Phys Chem B* 115**:**14327 −14332 (2011).

15. Ediger MD, Harrowell P (2012) Perspective: Supercooled liquids and glasses. *J. Chem. Phys.* 137**:**080901.

16. Kearns KL, Swallen SF, Ediger MD, Wu T, Yu L (2007) Influence of substrate temperature on the stability of glasses prepared by vapor deposition. *J Chem Phys* 127: 154702.

17. Leon-Gutierrez E, Sepúlveda A, Garcia G, Clavaguera-Mora MT, Rodríguez-Viejo J (2010) Stability of thin film glasses of toluene and ethylbenzene formed by vapor deposition: an in situ nanocalorimetric study. *Phys Chem Chem Phys* 12: 14693−14698.

18. Stillinger FH (1995) A topographic view of supercooled liquids and glass formation. *Science* 267**:**1935–1939.

19. Debenedetti PG, Stillinger FH (2001) Supercooled liquids and the glass transition. *Nature* 410:259−267.

20. Kearns KL, et al. (2008) Hiking down the energy landscape: progress toward the Kauzmann temperature via vapor deposition. *J Phys Chem B* 112:4934–4942.

21. Dawson KJ, et al. (2009) Highly Stable Indomethacin Glasses Resist Uptake of Water Vapor. *J Phys Chem B* 113:2422–2427.

22. Sokolov AP, Rössler E, Kisliuk A, Quitmann D (1993) Dynamics of strong and fragile glass formers: Differences and correlation with low-temperature properties. *Phys Rev Lett* 71:2062–2065.

23. Hédoux A, Guinet Y, Capet F, Paccou L, Descamps M (2008) Evidence for a high-density amorphous form in indomethacin from Raman scattering investigations. *Phys Rev B* 77:094205.

24. Kearns KL, Still T, Fytas G, Ediger MD (2010) High-Modulus Organic Glasses Prepared by Physical Vapor Deposition. *Adv Mater* 22:39–42.

25. Dawson K, Zhu L, Yu LA, Ediger MD (2011) Anisotropic Structure and Transformation Kinetics of Vapor-Deposited Indomethacin Glasses. *J Phys Chem B* 115:455−463.

26. Dalal SS, Ediger MD (2012) Molecular Orientation in Stable Glasses of Indomethacin. *J Phys Chem Lett* 3:1229−1233.

27. Singh S, De Pablo JJ (2011) A molecular view of vapor deposited glasses. *J Chem Phys* 134:194903.

28. Dawson K, et al. (2012) Molecular packing in highly stable glasses of vapor-deposited tris-naphthylbenzene isomers. *J Chem Phys* 136:094505.

29. Yokoyama DJ (2011) Molecular orientation in small-molecule organic light-emitting diodes. *J Mater Chem* 21:19187–19202.

30. Liu X, et al. (1997) Amorphous Solid without Low Energy Excitations. *Phys Rev Lett* 78:4418–4421.

31. Zink BL, Pietri R, Hellman F (2006) Thermal Conductivity and Specific Heat of Thin-Film Amorphous Silicon. *Phys Rev Lett* 96:055902.

32. Queen DR, Liu X, Karel J, Metcalf TH, Hellman F (2013) Excess Specific Heat in Evaporated Amorphous Silicon. *Phys Rev Lett* 110:135901.

33. Agladze NI, Sievers AJ (1998) Absence of an Isotope Effect in the Two Level Spectrum of Amorphous Ice. *Phys Rev Lett* 80:4209–4212.

34. Angell CA, Moynihan, Hemmati M (2000) 'Strong' and 'superstrong' liquids, and an approach to the perfect glass state via phase transition. *J Non-Cryst Solids* 274:319–331.

35. Ammann-Winkel K, et al. (2013) Water's second glass transition. *Proc Natl Acad Sci USA* 110(44):17720–17725.

36. Pérez-Castañeda T, Jiménez-Riobóo RJ, Ramos MA (2013) Low-temperature thermal properties of a hyperaged geological glass. *J Phys: Cond. Matter* 25:295402.

37. Pérez-Castañeda T, Jiménez-Riobóo RJ, Ramos MA (2014) Two-Level Systems and Boson Peak Remain Stable in 110-Million-Year-Old Amber Glass. *Phys Rev Lett* 112:165901.

38. Leggett AJ, Vural DC (2013) "Tunneling Two-Level Systems" Model of the Low-Temperature Properties of Glasses: Are "Smoking-Gun" Tests Possible? *J Phys Chem B* 117:12966–12971.

## Author contributions

J.R.V. and M.A.R. designed research; T.P.C. and C.R.T. performed research; T.P.C. and C.R.T. analyzed data; M.A.R. and J.R.V. wrote the paper.

## Supporting Information

### Specific-heat measurements at low temperatures

The low-temperature specific heat was measured in the temperature range $0.18 \text{ K} \leq T \leq$ 32 K by means of thermal relaxation calorimetry, with the experimental setup described in more detail elsewhere (1,2), employing both a [4]He cryostat (above 2 K) and a [3]He-[4]He dilution refrigerator (below 3 K), to cover a temperature range $0.18 \text{ K} \leq T \leq 32$ K. Two slightly different methods were used depending on the relaxation time constants. For values of the order of one minute, the standard relaxation method (see Fig. S1) was employed. For higher relaxation times at higher temperatures, an alternative faster relaxation method was chosen (3), in which the heating power is switched off before reaching the equilibrium. Combining the thermal charging or heating step (where the thermal conductance $K$ between the cell and the thermal bath is determined) and the relaxation step (where the relaxation time $\tau$ is directly obtained), the specific heat is straightforward calculated from $C_p = K \cdot \tau$ (where $K = P/\Delta T_\infty$, $P$ being the applied power and $\Delta T_\infty$ the corresponding thermal jump) in both cases (3).

The accurate determination of the specific heat requires high thermal stability, given by thermal drifts of the order of few hundredths to few tenths of µK/s. Also, temperature jumps systematically smaller than $\Delta T/T_0 \leq 2\%$ where applied, so that the thermal conductance of the cell-to-surroundings contact could be approximated as constant in the heating-relaxation process. The insets in Fig. S1 show the logarithm decrease of temperature versus time during the relaxation. The linearity of this semilogarithmic plot confirms the suitability of the method, dominated by a single relaxation time (the thermal link) via $T(t) = T_0(t) + \Delta T_\infty \cdot \exp(-t/\tau)$, even at the lowest temperatures where the thermal stability is more difficult (see lower panel of Fig. S1).

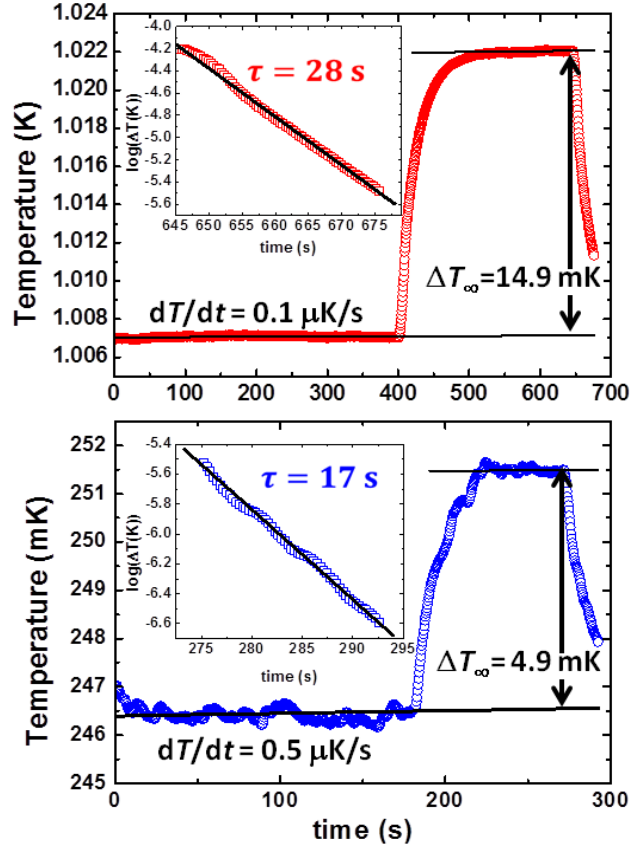

**Fig. S1.** Examples of experimental points at 1.0 K (upper panel) and 0.25 K (lower panel), using the standard thermal relaxation method. After establishing the thermal background, a constant power is applied to the calorimetric cell with the sample. Once the temperature drift is parallel to the previous background, the power is switched off and the heat is exponentially released through the thermal link. The corresponding relaxation time $\tau$ is straightforwardly obtained from a semilogarithmic plot (insets), from which the heat capacity is determined.

Alternatively, when the heat capacity to be measured is larger (either at higher temperatures or when the sample mass is larger) and hence the relaxation time becomes significantly longer, a faster relaxation procedure (3) was followed (see an example in Fig. S2). In brief, instead of waiting until a full thermal equilibration after applying the power, the unknown long-time thermal increase $\Delta T_\infty$ is determined (3) from the exponential heating curve, $T(t) = T_0(t) + \Delta T_\infty \cdot [1 - \exp(-t/\tau)]$, where $\tau$ is obtained from the corresponding relaxation cooling curve, as in the standard method.

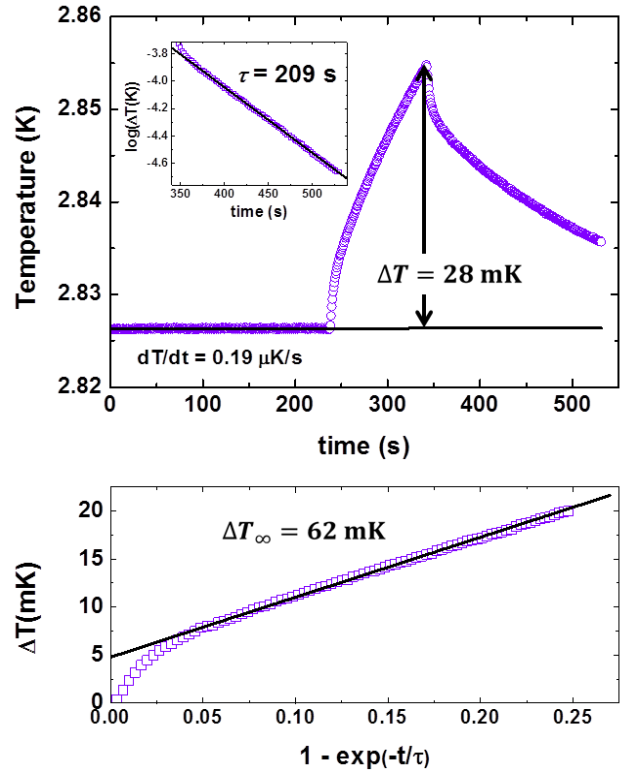

**Fig. S2.** Example of an experimental point at about 2.8 K, for a sample with a higher heat capacity than in Fig. S1, using the alternative thermal relaxation method. Upper panel: After establishing the thermal background, a constant power is applied to the calorimetric cell with the sample. Without waiting until the thermal equilibrium is achieved, the power is switched off and the heat is exponentially released through the thermal link. The corresponding relaxation time $\tau$ is obtained from a semilogarithmic plot (inset). In this alternative method, the needed factor $\Delta T_\infty$ is determined from a linear fit of the heating curve as shown in the lower panel.

## Supporting references

1. Pérez-Castañeda T, et al. (2013) Low-temperature specific heat of graphite and CeSb₂: Validation of a quasi-adiabatic continuous method. *J Low Temp Phys* 173:4−20.

2. Pérez-Castañeda T, Jiménez-Riobóo RJ, Ramos MA (2013) Low-temperature thermal properties of a hyperaged geological glass. *J Phys: Condens Matter* 25:295402.

3. Pérez-Enciso E, Ramos MA (2007) Low-temperature calorimetry on molecular glasses and crystals. *Thermochimica Acta* 461:50−56.